\documentclass[11pt]{article}
\usepackage{amsmath,amssymb}
\usepackage{graphicx}
\usepackage{subcaption}
\usepackage{xcolor}

\begin{document}

\setcounter{page}{1}

\pagestyle{plain}

\begin{center}
\Large{\bf Observational Viability of $\phi^{2}$-Superpotential Inflation with GUP-Induced Corrections}\\
\small \vspace{1cm} {Narges
Rashidi}\footnote{n.rashidi@umz.ac.ir (Corresponding Author)} and {Maryam Roushan}\footnote{m.roushan@umz.ac.ir} \\
\vspace{0.5cm} $^{}$ Department of Theoretical Physics, Faculty of
Science,
University of Mazandaran,\\
P. O. Box 47416-95447, Babolsar, IRAN\\
\end{center}

\begin{abstract}
We study a $\phi^{2}$-superpotential inflationary model within a GUP-inspired quantum-gravity framework. Using horizon thermodynamics, we review the GUP-corrected Friedmann equations obtained by combining the temperature--surface-gravity relation with a modified entropy--area law. Reformulating the dynamics via a superpotential $W(\phi)$ with $\phi=\phi(a)$, we obtain the corresponding slow-roll parameters and derive GUP-modified expressions for the scalar spectral index and the tensor-to-scalar ratio, leading to a deformation of the standard inflationary consistency relation. Adopting the quadratic ansatz $W(\phi)=\tfrac{1}{2}m^{2}\phi^{2}$, we evaluate the observables $(n_{s},r)$ and compare them with the Planck 2018 TT, TE, EE + lowE + lensing + BK18 + BAO, DESI + CMB + DESY5, and Planck2018 + ACT + lensing + BK18 + BAO (from DESI) datasets. The analysis shows that, for moderately negative values of the effective GUP parameter $\beta$ and sub-unity values of the phenomenological superpotential parameter $C$, the model predictions can fall within the $68\%$ and $95\%$ confidence regions of current observations. These results indicate that GUP-inspired corrections can shift the $(n_{s},r)$ predictions of a quadratic superpotential toward the observationally favored region.
\end{abstract}
\newpage

\section{Introduction}
\label{sec1}
In the framework of relativistic cosmology, the behavior of the
universe on large scales is usually described through the Friedmann
equations, which stem from the Einstein field equations. These
equations play a central role in describing the expansion history of
the universe. Jacobson argued in~\cite{Jac95} that one can, in fact,
rederive the Einstein equations by adopting a thermodynamic
viewpoint. His approach relies on the Clausius relation,
$dQ = T\,dS$, together with the entropy--area correspondence known
from black-hole thermodynamics, and assumes that such a relation is
valid for all local Rindler horizons. Within this interpretation,
the Einstein field equations appear as thermodynamic equations of
state for spacetime itself (see also~\cite{Jac95}). Later,
Verlinde~\cite{Ver00} pointed out that the Friedmann relation in a
FRW background can be written in a form similar to the entropy law
governing the thermodynamics of the radiation component. Related
developments and discussions in this direction can be found
in~\cite{Cai03,Noj01,Noj02,Cve02}. These studies collectively
highlight a close connection between thermodynamics and
gravitational dynamics, motivating further investigations in
different contexts. For example, Cai and
collaborators~\cite{Cai05} derived the Friedmann equations in
$(n+1)$ dimensions from a purely thermodynamic analysis.

A further line of research concerns the role of the uncertainty
principle in the phenomenon of Hawking radiation within black-hole
physics~\cite{Med04,Adl01,Cav03,Cav04,Maj11,Ali12a,Ali12b}. In this
setting, the so-called Generalized Uncertainty Principle (GUP),
emerging from considerations in quantum gravity and string theory,
has become especially significant. The GUP introduces a deformation
of the standard Heisenberg uncertainty relation, which in many
realizations is associated with the emergence of an effective
minimal length scale. The modified form can be expressed
as~\cite{Cav03}
\begin{equation}
	\label{eq1}
	\Delta x_{i}\,\Delta
	p_{i} \gtrsim \left[\hbar+\frac{\beta\, l_{Pl}^{2}}{\hbar}
	\left(\Delta p_{i}\right)^{2}\right]\,,
\end{equation}
where $l_{Pl}$ denotes the Planck length and $\beta$ is a
dimensionless deformation parameter whose sign and magnitude depend
on the underlying effective framework. {Note that within the generalized uncertainty principle (GUP) framework, $\beta$ characterizes the strength of the quantum-gravity-induced deformation of the standard Heisenberg uncertainty relation. In many minimal-length realizations of the GUP, $\beta$ is usually assumed to be positive. However, its sign is not universally fixed and depends on the underlying effective quantum-gravity framework~\cite{Scardigli2019}. Negative values of $\beta$ have also been discussed in several independent contexts, including lattice-based uncertainty relations~\cite{Jiz10}, analyses of the Chandrasekhar limit for white dwarfs~\cite{Ong2018}, black-hole thermodynamics~\cite{Buoninfante2019,Zhou2022}, and studies relating GUP corrections to corpuscular gravity~\cite{Buoninfante2019}. These investigations indicate that the negative-$\beta$ branch can possess a meaningful physical interpretation and should not be regarded merely as a mathematical possibility~\cite{Scardigli2019,Buoninfante2019,Zhou2022}.	The magnitude of $\beta$ is likewise not uniquely determined. For example, by comparing GUP-modified black-hole thermodynamics with the corpuscular gravity picture, Buoninfante \textit{et al.} estimated the deformation parameter to be of order unity, $|\beta|\sim \mathcal{O}(1)$~\cite{Buoninfante2019}. On the phenomenological side, Scardigli reviewed a variety of experimental and observational constraints obtained in different physical systems~\cite{Scardigli2019}, including upper bounds of approximately $\beta<10^{50}$ from Landau levels~\cite{Das08,Ali11}, $\beta<10^{36}$ from the Lamb shift~\cite{Das08,Ali11}, $\beta<10^{34}$ from charmonium spectroscopy~\cite{Das08,Ali11}, $\beta<10^{21}$ from equivalence-principle tests~\cite{Go14}, $\beta<10^{12}$~\cite{Ba15} and, in some analyses, $\beta<5\times10^{6}$~\cite{Bu19} from macroscopic oscillators, and $\beta<10^{60}$ from gravitational-wave observations~\cite{Fe17}. These bounds span many orders of magnitude and generally provide only upper limits on the magnitude of $\beta$, while no universal lower bound has been established. Therefore, no universal model-independent constraint on either the sign or magnitude of $\beta$ currently exists. In the present work, we therefore treat $\beta$ as an effective phenomenological parameter entering the entropy-area relation and the resulting GUP-corrected Friedmann equations. We do not impose an \emph{a priori} restriction on its sign. Instead, the viable range of $\beta$ is determined through comparison with cosmological observations. As will be shown below, the parameter regions compatible with current observational constraints correspond to moderately negative values of $\beta$. This preference emerges from the observational analysis of the model rather than being assumed from the outset. Equation~(\ref{eq1}),
together with the identification $\Delta p \sim 1/\Delta x$ and
hence $\Delta E \sim 1/\Delta x$~\cite{Cam04}, implies nontrivial
corrections to the energy scale associated with localized probe
particles. Since the entropy of a black hole is linked to its area,
and the area in turn is related to the energy, these corrections
modify the entropy--area relation. Incorporating the resulting
entropy correction into the first law of thermodynamics ultimately
yields a modified version of the Friedmann equation~\cite{Awa14}.

Recent developments in quantum-gravity phenomenology suggest that
GUP-inspired deformations can lead to nontrivial corrections in the
thermodynamic derivation of the Friedmann equations. These
corrections modify the background dynamics and deform the comoving
momentum of perturbation modes, resulting in potentially observable
shifts in the primordial spectra. Since inflation operates only a
few orders of magnitude below the Planck scale, examining how such
quantum-gravity-inspired corrections influence key observables such
as $n_s$ and $r$ becomes physically well motivated.

An essential application of the Friedmann equations arises in the
theory of primordial inflation. During this epoch, the universe
underwent an extremely rapid, nearly exponential expansion within a
tiny fraction of a second. Numerous scenarios have been developed to
explain this process. Some scenarios predict nearly Gaussian
primordial perturbations, while others can generate significant
non-Gaussian signatures
\cite{Gut81,Lin82,Alb82,Lin90,Lid00a,Lid97,Rio02,Lyt09,Mal03}.
Any inflationary proposal, however, must ultimately be assessed in
light of observations. This requires evaluating key perturbation
quantities such as the scalar spectral index $n_{s}$, the tensor
spectral index $n_{T}$, and the tensor-to-scalar ratio $r$.
Comparing these predictions with recent cosmological observations
provides an important test of the phenomenological viability of
inflationary models.

One of the simplest inflationary scenarios is the single-field model
with a quadratic potential, $V(\phi)\propto \phi^{2}$. Although this
setup satisfies the slow-roll conditions and provides sufficient
e-folds of inflation, its standard canonical realization is
disfavored by current observations. In particular, the predicted
values of $n_{s}$ and $r$ place the corresponding $r-n_{s}$ curve
outside the $68\%$ and, in significant regions, the $95\%$
confidence contours of the Planck 2018 results
\cite{pl18b,pa22}, as well as more recent DESI constraints
\cite{wa24,co24}. These tensions have motivated the exploration of
inflationary scenarios beyond the minimal quadratic framework,
either through modifications of the gravitational sector or through
more generalized realizations of scalar-field dynamics.

In this regard, cosmologists have considered a variety of extensions of
the simplest inflationary models. For instance, Hilltop quartic
inflation can be consistent with observational data in certain regions
of its parameter space \cite{pl18b}, while $\alpha$-attractor models
admit domains in which inflationary observables fall within current
constraints \cite{pl18b}. Another theoretically well-motivated
framework is based on the superpotential formalism, in which the scalar
potential is generated dynamically from a first-order structure. In
this generalized superpotential framework, we consider the inflationary
potential in the form
\begin{equation} \label{eq2}
	V(\phi)=\frac{C^{2}}{\kappa^{4}}
	\bigg(3\,\kappa^{2}\,W(\phi)^{2}
	-2\,W_{,\phi}(\phi)^{2}\bigg)\,.
\end{equation}
Here $W(\phi)$ denotes a general function of the scalar field, while $C$ is a dimensionless parameter characterizing the generalized reconstruction of the scalar potential within the superpotential framework. It is important to emphasize that no assumption
of supersymmetry has been made in this context. Nevertheless, because
the structure of the potential in (\ref{eq2}) resembles that encountered
in supergravity frameworks, it is conventionally referred to as the
superpotential. Crucially, even when a quadratic functional form is
adopted for $W(\phi)$, the resulting inflationary dynamics are not
equivalent to those of the canonical quadratic potential minimally
coupled to General Relativity. The effective scalar potential is
generated nonlinearly through the Hamilton--Jacobi formulation,
leading to background evolution and consistency relations that are
structurally distinct from standard quadratic inflation. In this work,
we therefore employ the superpotential framework as a controlled
theoretical benchmark to investigate how inflationary predictions
depend on modified dynamical relations and possible quantum-gravity
effects, rather than to restore the phenomenological viability of an
otherwise disfavored canonical potential. This viewpoint is consistent
with several recent studies in which quadratic scalar structures are
deliberately revisited as controlled theoretical benchmarks within
generalized inflationary frameworks, rather than as phenomenologically
preferred models, see e.g. \cite{Odintsov2023,Wang2024}. In the present
framework, the Hubble expansion rate is generated through a generalized
Hamilton--Jacobi relation involving the superpotential function. This
formulation allows the potential to be generated self-consistently from
the dynamics. When GUP-inspired corrections introduce nonlinear
modifications to the Friedmann equations, this first-order structure
provides a convenient framework for incorporating the corrections into
the inflationary dynamics without introducing higher-derivative terms
at the effective level.

Combining the GUP-inspired corrected background dynamics with the
superpotential framework enables us to address a concrete and
physically relevant question: \emph{Can GUP-induced corrections shift
the predictions of a quadratic superpotential model toward agreement
with modern cosmological observations?} As we show, for certain
ranges of the model parameters, the predicted values of the scalar
spectral index and tensor-to-scalar ratio fall within the
$68\%-95\%$ confidence regions of the Planck, DESI, and ACT datasets,
demonstrating that the interplay between GUP-inspired corrections and
superpotential inflation can lead to observationally viable
predictions.

The remainder of this paper is organized as follows.
In Section~\ref{sec2}, we derive the GUP-corrected Friedmann
equations from a thermodynamic perspective.
Section~\ref{sec3} is devoted to studying the superpotential-based
inflationary model within this framework.
In Section~\ref{sec4}, we present a numerical analysis of the model
and confront its predictions with observational data.

\section{GUP-Corrected Friedmann Equations from a Thermodynamic Perspective
	\label{sec2}}

To derive the modified form of the Friedmann equations, we begin with
a general FRW geometry in $(3+1)$ dimensions, whose metric can
be written as
\begin{eqnarray}
	\label{eq3}
	ds^2=h_{ij}\,dx^i\,dx^{j}+{\cal{X}}^{2}\,d\Omega_{2}^{2}\,,
\end{eqnarray}
with the two-dimensional metric
$h_{ij}=\text{diag}(-1,\,a^{2}/(1-kr^2))$, the areal radius
${\cal{X}}=a(t)\,r$, and coordinates $x^{i}=(t,r)$. Here, $k=0,\pm 1$ specifies the spatial curvature and
$d\Omega_2^{2}$ denotes the metric of a unit $2$-sphere.
In the following, we focus on the spatially flat case, $k=0$, which is
the background relevant for the inflationary analysis considered in
this work. For this flat FRW background, the apparent horizon is
\cite{Cai05}
\begin{eqnarray}
	\label{eq4}
	{\cal{X}}_{A}=a\,r=\frac{1}{H}\,,
\end{eqnarray}
where the Hubble parameter is $H=\dot{a}/a$, with the dot denoting a
derivative with respect to cosmic time. In FRW spacetime the thermodynamically relevant surface is the apparent horizon rather than the event horizon. The apparent horizon admits a well-defined Kodama vector and associated surface gravity, which allows one to formulate the unified first law,
	\begin{eqnarray}
		\label{eq5} dE=T\,dS+{\cal{W}}\,dV\,,
	\end{eqnarray}
	with $E$ denoting the total energy enclosed by the apparent horizon. This framework was established by Hayward~\cite{Ha99,Hay98}. Our derivation follows the same apparent-horizon construction. To proceed, we evaluate each term appearing in Eq.~(\ref{eq5}). In Ref.~\cite{Awa14}, the link between the horizon temperature and
the surface gravity is expressed as
\begin{eqnarray}
	\label{eq6}
	T=\frac{\kappa}{2\pi}\,,
\end{eqnarray}
where the quantity $\kappa$ denotes the surface gravity, given by
\begin{eqnarray}
	\label{eq7}
	\kappa=-\frac{1}{{\cal{X}}_{A}}+\frac{\dot{{\cal{X}}}_{A}}{2H{\cal{X}}_{A}^{2}}
\end{eqnarray}
in the FRW background. Note that the Gibbons--Hawking effect may be interpreted as a squeezed quantum vacuum rather than a strictly thermal bath. Nevertheless, the quantity entering the thermodynamic derivation is the effective temperature associated with the surface gravity of the Kodama observer. 

Furthermore, following Ref.~\cite{Awa14}, the entropy associated with
the horizon area $A$ is parameterized as
\begin{eqnarray}
	\label{eq8}
	S=\frac{f(A)}{4G}\,.
\end{eqnarray}
Combining Eqs.~(\ref{eq6})--(\ref{eq8}) yields
\begin{eqnarray}
	\label{eq9}
	T\,dS=
	\left(
	-\frac{1}{2\pi{\cal{X}}_{A}}
	+\frac{\dot{{\cal{X}}}_{A}}
	{4\pi H{\cal{X}}_{A}^{2}}
	\right)
	\left(
	\frac{df(A)}{dA}
	\right)
	\left(
	\frac{8\pi{\cal{X}}_{A}}
	{4G}\,d{\cal{X}}_{A}
	\right)\,.
\end{eqnarray}

To describe the matter content, we assume a perfect fluid satisfying
the conservation equation
$\dot{\rho}+3H(\rho+p)=0$, where $\rho$ and $p$ denote the energy
density and pressure, respectively. For a spatially flat FRW
background containing a perfect fluid, one obtains the following
expression for the work function~\cite{Hay98,Bak00}
\begin{eqnarray}
	\label{eq10}
	{\cal{W}}=\frac{1}{2}(\rho-p)\,.
\end{eqnarray}

The Misner--Sharp energy, which represents the total energy enclosed
by the apparent horizon, can be written as
$E=\rho V$~\cite{Awa14,ch11}, where $V$ denotes the volume enclosed
by the apparent horizon,
$V=\frac{4}{3}\pi {\cal{X}}_{A}^{3}$. Using these definitions, one
obtains
\begin{eqnarray}
	\label{eq11}
	dE=
	4\pi{\cal{X}}_{A}^{2}\,\rho\,d{\cal{X}}_{A}
	-
	4\pi\,{\cal{X}}_{A}^{3}
	(\rho+p)\,H\,dt\,.
\end{eqnarray}

Similarly,
\begin{eqnarray}
	\label{eq12}
	{\cal{W}}\,dV=
	2\pi\,{\cal{X}}_{A}^{2}
	(\rho-p)\,d{\cal{X}}_{A}\,.
\end{eqnarray}

Combining Eqs.~(\ref{eq9})--(\ref{eq12}), one obtains the following
form of the second Friedmann equation
\begin{eqnarray}
	\label{eq13}
	\dot{H}\,f'(A)
	=
	-4\pi\,G\,(\rho+p)\,,
\end{eqnarray}
where a prime denotes differentiation with respect to $A$.
Integrating the above expression together with the conservation
equation yields the corresponding Friedmann equation
\begin{eqnarray}
	\label{eq14}
	\rho=
	-\frac{3}{2\,G}
	\int
	\frac{f'(A)}{A^2}\,dA\,.
\end{eqnarray}

Up to this point, the derivation has relied purely on classical
thermodynamics. To incorporate quantum-gravity-inspired effects, we
now extend the analysis by including the generalized uncertainty
principle (GUP), which introduces an effective deformation of the
standard uncertainty relation into the thermodynamic description of
spacetime. At this stage, it is worth clarifying the conceptual
framework adopted in the present analysis. The GUP corrections are
implemented through horizon thermodynamics, rather than as direct
modifications of relativistic quantum-field dynamics. Horizon
thermodynamics provides an effective reformulation of Einstein
gravity, in which corrections to the entropy--area relation
naturally translate into modified cosmological dynamics while
preserving the relativistic structure of the theory. Similar
treatments of uncertainty-principle--induced corrections in
relativistic and cosmological settings have been explored in the
literature, including studies within EPJC and related journals
\cite{Dabrowski2019,Dabrowski2020,Luciano2021,Giardino2021,Lidsey2013}.

When a black hole absorbs or emits a quantum particle, its area
changes by at least
$\Delta A\geq 8\pi l_{Pl}^{2}ER$, with $R$ and $E$ denoting the
particle size and energy, respectively~\cite{Chr70}. Assuming that
the characteristic particle size is associated with the position
uncertainty, $R\simeq\Delta x$, one obtains
$\Delta A_{\min}\geq 8\pi l_{Pl}^{2}E\Delta x$. For positive values
of the deformation parameter $\beta$, Eq.~(\ref{eq1}) formally
implies the existence of an effective minimal position uncertainty
$\Delta x_{\min}=2\sqrt{\beta}\,l_{Pl}$~\cite{Awa14}. More generally,
in the present work we treat $\beta$ as an effective GUP deformation
parameter entering the thermodynamic description of spacetime.
Using the relation
$E\geq1/\Delta x$~\cite{Med04}, together with
Eq.~(\ref{eq1}), and identifying
$\Delta x^{2}=A/\pi$, the area variation can be written
as~\cite{Awa14}
\begin{eqnarray}
	\label{eq15}
	\Delta A_{\min}\simeq
	\chi\,
	\frac{8A}{2\beta}
	\left(
	1-\sqrt{1-\frac{4\beta\pi l_{Pl}^{2}}{A}}
	\right).
\end{eqnarray}

The constant $\chi$ is fixed by the Bekenstein--Hawking entropy.
Since the minimal entropy increment is a single bit,
$\Delta S_{\min}=\ln2$, and using the relation
$b/\chi=2\pi$~\cite{Med04}, one obtains
\begin{eqnarray}
	\label{eq16}
	\frac{dS}{dA}
	=
	\frac{\Delta S_{\min}}{\Delta A_{\min}}
	=
	\frac{\pi}{
		\tfrac{2A}{\beta}
		\left(
		1-\sqrt{
			1-\tfrac{4\beta\pi l_{Pl}^{2}}{A}}
		\right)}.
\end{eqnarray}

Starting from Eq.~(\ref{eq8}), one obtains the relation
$f'(A)=4G\,dS/dA$. Substituting the entropy--area expression
(\ref{eq16}) into this relation allows the Friedmann equations
(\ref{eq13}) and (\ref{eq14}) to be rewritten in the form
\begin{eqnarray}
	\label{eq17}
	\dot{H}
	\left(
	\frac{
		\beta\,l_{Pl}^{2}\,H^{2}}
	{2-2\sqrt{1-\beta\,l_{Pl}^{2}H^{2}}}
	\right)
	=
	-4\pi G(\rho+p)\,,
\end{eqnarray}
\begin{eqnarray}
	\label{eq18}
	\frac{H^{2}}{2}
	+
	\frac{
		\left[
		1-
		\left(
		1-\beta\,l_{Pl}^{2}H^{2}
		\right)^{3/2}
		\right]}
	{3\beta\,l_{Pl}^{2}}
	=
	\frac{8\pi G}{3}\rho\,.
\end{eqnarray}

Here we have used the area--radius relation
$A=4\pi{\cal{X}}_{A}^{2}$. In this framework, the GUP corrections
enter the gravitational dynamics through the modified entropy--area
relation within horizon thermodynamics. Substituting the corrected
expression for $dS/dA$ into the unified first law at the apparent
horizon directly yields the modified Friedmann equations.
Equations~(\ref{eq17}) and~(\ref{eq18}) therefore represent the
GUP-corrected Friedmann equations that will serve as the background
dynamics for the inflationary analysis presented in the next section.

\section{Superpotential-Based Inflation with GUP Corrections
	\label{sec3}}

To analyze the inflationary scenario driven by the superpotential
framework, we express the inflaton field as a function of the scale
factor, $\phi=\phi(a)$. This allows the GUP-corrected Friedmann
equations derived in the previous section to be reformulated in terms
of the superpotential and its derivatives. In this section, a prime
denotes differentiation with respect to $\phi$. Proceeding in this
way, one obtains the following rewritten forms of
Eqs.~(\ref{eq17}) and~(\ref{eq18}):
\begin{eqnarray}
	\label{eq19}
	\left(
	H\,H'\,
	a\,\frac{d\phi}{da}
	\right)
	\left(
	\frac{
		\beta\,l_{Pl}^{2}\,H^{2}}
	{2-2\sqrt{1-\beta\,l_{Pl}^{2}H^{2}}}
	\right)
	=
	-4\pi G\,H^{2}
	\left(
	a\,\frac{d\phi}{da}
	\right)^{2}\,,
\end{eqnarray}

\begin{eqnarray}
	\label{eq20}
	\frac{H^{2}}{2}
	+
	\frac{
		\left[
		1-
		\left(
		1-\beta\,l_{Pl}^{2}H^{2}
		\right)^{3/2}
		\right]}
	{3\beta\,l_{Pl}^{2}}
	=
	\frac{8\pi G}{3}
	\left[
	\frac{1}{2}
	H^{2}
	\left(
	a\,\frac{d\phi}{da}
	\right)^{2}
	+
	V(\phi)
	\right]\,.
\end{eqnarray}

The corresponding equation of motion for the inflaton field takes
the form
\begin{equation}
	\label{eq21}
	\ddot{a}\,\frac{d\phi}{da}
	+
	\dot{a}^{2}\,\frac{d^{2}\phi}{da^{2}}
	+
	3H^{2}a\,\frac{d\phi}{da}
	+
	V'(\phi)
	=
	0\,.
\end{equation}
Throughout this work, we adopt natural units
$c=\hbar=1$. In these units, the superpotential
$W(\phi)$ carries the same mass dimension as the
Hubble parameter.
Within the Hamilton--Jacobi formulation adopted in the present work,
the Hubble parameter is identified with the superpotential function as
\begin{eqnarray}
	\label{eq22}
	H(\phi)=W(\phi)\,.
\end{eqnarray}

Using Eqs.~(\ref{eq2}), (\ref{eq20}), and (\ref{eq22}), and defining
$X\equiv a\,d\phi/da$, one obtains
\begin{eqnarray}
	\label{eq23}
	a\,\frac{d\phi}{da}
	=
	\pm
	\Bigg[
	\frac{3}{\kappa^{2}}
	+
	\frac{
		2\left[
		1-\left(1-\beta l_{Pl}^{2}W^{2}\right)^{3/2}
		\right]}
	{\beta l_{Pl}^{2}\kappa^{2}W^{2}}
	-
	\frac{2C^{2}}{\kappa^{4}W^{2}}
	\left(
	3\kappa^{2}W^{2}-2W_{,\phi}^{2}
	\right)
	\Bigg]^{1/2}.
\end{eqnarray}

In the numerical analysis presented below, we select the branch
corresponding to a monotonically decreasing inflaton field,
$\dot{\phi}<0$, which implies
$a\,d\phi/da<0$ during the inflationary expansion.

At this stage, the slow-roll parameters of the model can be
evaluated. From the definitions
$\epsilon=-\dot{H}/H^{2}$ and
$\eta=H^{-1}\dot{\epsilon}/\epsilon$, and using
$H(\phi)=W(\phi)$, one obtains
\begin{eqnarray}
	\label{eq24}
	\epsilon
	=
	-\frac{W_{,\phi}}{W}
	a\,\frac{d\phi}{da}\,.
\end{eqnarray}
It is useful to define the function
\begin{eqnarray}
	\label{eq25a}
	{\cal F}(\phi)
	\equiv
	\frac{3}{\kappa^{2}}
	+
	\frac{
		2\left[
		1-\left(1-\beta l_{Pl}^{2}W^{2}\right)^{3/2}
		\right]}
	{\beta l_{Pl}^{2}\kappa^{2}W^{2}}
	-
	\frac{2C^{2}}{\kappa^{4}W^{2}}
	\left(
	3\kappa^{2}W^{2}-2W_{,\phi}^{2}
	\right).
\end{eqnarray}
According to Eq.~(\ref{eq23}), one has
$a\,d\phi/da=\pm\sqrt{{\cal F}(\phi)}$. Therefore, the first
slow-roll parameter can be written as
\begin{eqnarray}
	\label{eq24b}
	\epsilon
	=
	\mp
	\frac{W_{,\phi}}{W}
	\sqrt{{\cal F}(\phi)}\,.
\end{eqnarray}
The sign is chosen according to the direction of the inflaton
evolution.

Similarly, using
$\eta=(\epsilon_{,\phi}/\epsilon)\,a\,d\phi/da$, we find
\begin{eqnarray}
	\label{eq25}
	\eta
	=
	\pm
	\sqrt{{\cal F}(\phi)}
	\left[
	\frac{W_{,\phi\phi}}{W_{,\phi}}
	-
	\frac{W_{,\phi}}{W}
	+
	\frac{1}{2}
	\frac{{\cal F}_{,\phi}}{{\cal F}}
	\right].
\end{eqnarray}
We next turn to the analysis of perturbation parameters in this
framework in order to clarify how GUP-inspired corrections affect
the inflationary dynamics. In this context, the comoving wave number
plays a crucial role, since its effective form is modified once the
generalized uncertainty principle is incorporated. At the effective
level, the generalized uncertainty relation (\ref{eq1}) can be
associated with a modified commutation relation of the form
~\cite{Med04,Awa14}
\begin{eqnarray}
	\label{eq26}
	[X_{i},P_{j}]
	=
	i\Big(
	\delta_{ij}
	+
	\beta_{ijkl}\,
	l_{Pl}^{2}\,
	p^{k}p^{l}
	\Big)\,.
\end{eqnarray}
A commonly used realization of the modified algebra is given by
\begin{eqnarray}
	\label{eq27}
	X^{i}=x^{i}\,,
	\qquad
	P^{i}=p^{i}
	\left(
	1+\beta\,l_{Pl}^{2}\,p^{2}
	\right)\,.
\end{eqnarray}

From Eq.~(\ref{eq27}), together with the identification $p=k$, one
obtains the effective modified wave number~\cite{Ras23}
\begin{eqnarray}
	\label{eq28}
	K^{i}
	=
	k^{i}
	\left(
	1+\beta\,l_{Pl}^{2}\,k^{2}
	\right)\,,
\end{eqnarray}
which leads to the effective UV-modified dispersion relation
\[
\omega^{2}=K^{2}.
\]

In this framework, the modified wave number induces corrections to
the inflationary perturbation parameters. In the present work, we
adopt the standard effective GUP-based implementation of
UV-modified dispersion relations commonly used in the literature
\cite{Tw13,Ras23}.

The scalar spectral index is then defined as
\cite{Tw13,Ras23}
\begin{eqnarray}
	\label{eq29}
	n_{s}
	=
	1+
	\frac{
		d\ln{\cal A}_{s}}
	{d\ln\!\left[
		k\left(
		1+\beta\,l_{Pl}^{2}\,k^{2}
		\right)
		\right]}
	=
	1+
	\left(
	1+\beta\,l_{Pl}^{2}\,k^{2}
	\right)
	(-2\epsilon-\eta)\,,
\end{eqnarray}
with
\begin{equation}
	{\cal A}_{s}
	=
	\frac{H^{2}}
	{8\pi^{2}\epsilon}\,.
\end{equation}

Similarly, the tensor spectral index becomes
\begin{eqnarray}
	\label{eq30}
	n_{T}
	=
	\frac{
		d\ln{\cal A}_{T}}
	{d\ln\!\left[
		k\left(
		1+\beta\,l_{Pl}^{2}\,k^{2}
		\right)
		\right]}
	=
	-
	\left(
	1+\beta\,l_{Pl}^{2}\,k^{2}
	\right)
	(2\epsilon)\,,
\end{eqnarray}
with
\begin{equation}
	{\cal A}_{T}
	=
	\frac{2H^{2}}
	{\pi^{2}}\,.
\end{equation}

In the numerical analysis presented below, all inflationary
observables are evaluated at the pivot scale
$k_{*}=0.05\,{\rm Mpc}^{-1}$.

The resulting expressions for $n_s$ and $n_T$ therefore encode the
leading GUP-induced corrections without requiring a complete
re-derivation of the Mukhanov--Sasaki equation with an explicit
modified dispersion relation. Within the effective framework adopted
here, this prescription captures the dominant corrections while
smoothly reproducing the standard slow-roll limit as
$\beta\to0$.

The tensor-to-scalar ratio in the present setup becomes
\begin{equation}
	\label{eq31}
	r=
	\frac{{\cal A}_{T}}{{\cal A}_{s}}
	=
	16\epsilon
	=
	-\frac{
		8}
	{1+\beta l_{Pl}^{2}k^{2}}
	n_{T}\,.
\end{equation}

It follows that the standard consistency relation is modified in
this framework due to the GUP-induced corrections. Note that the
slow-roll parameters appearing in
Eqs.~(\ref{eq29})--(\ref{eq31}) are determined by
Eqs.~(\ref{eq24b}) and~(\ref{eq25}). Since the GUP modification alters the Friedmann equations, the
background quantities $H(\phi)$, $H_{,\phi}$, and $\dot{\phi}$
already contain the corresponding GUP corrections. These corrected
background quantities are then used consistently in the effective
perturbation sector.

At this stage, the viability of the model can be examined against
observational constraints. In what follows, we adopt the quadratic
superpotential ansatz
\[
W(\phi)=\frac{1}{2}m^{2}\phi^{2}.
\]
Although simple quadratic inflationary scenarios are disfavored in
the context of standard single-field inflation, our aim here is to
investigate whether embedding a quadratic superpotential within the
generalized superpotential framework together with GUP-induced
corrections can shift the inflationary predictions toward
observationally viable regions.

\section{Observational Analysis
	\label{sec4}}

In this section, we confront the predictions of the
$\phi^{2}$-superpotential inflationary model with recent
cosmological observations. Using the GUP-corrected Friedmann
equations and slow-roll relations derived in the previous sections,
we evaluate the main inflationary observables, namely the scalar
spectral index $n_s$ and the tensor-to-scalar ratio $r$, and compare
the results with the latest \textit{Planck}, \textit{DESI}, and
\textit{ACT} datasets.

For the quadratic superpotential ansatz
\[
W(\phi)=\frac{1}{2}m^{2}\phi^{2},
\]
we obtain
\begin{eqnarray}
	\label{eq32a}
	{\cal F}(\phi)
	=
	\frac{3}{\kappa^{2}}
	+
	\frac{
		2\left[
		1-
		\left(
		1-\frac14\beta l_{Pl}^{2}m^{4}\phi^{4}
		\right)^{3/2}
		\right]}
	{
		\frac14\beta l_{Pl}^{2}\kappa^{2}m^{4}\phi^{4}
	}
	-
	\frac{6C^{2}}{\kappa^{2}}
	+
	\frac{16C^{2}}{\kappa^{4}\phi^{2}}.
\end{eqnarray}
The slow-roll parameters are then given by
\begin{eqnarray}
	\label{eq32}
	\epsilon
	=
	\mp
	\frac{2}{\phi}
	\sqrt{{\cal F}(\phi)},
\end{eqnarray}
and
\begin{eqnarray}
	\label{eq33}
	\eta
	=
	\pm
	\sqrt{{\cal F}(\phi)}
	\left[
	-\frac{1}{\phi}
	+
	\frac{1}{2}
	\frac{{\cal F}_{,\phi}}
	{{\cal F}}
	\right].
\end{eqnarray}

To evaluate the inflationary observables, the scalar field must be
expressed in terms of the number of e-folds $N$ and the model
parameters. Starting from the definition
$N=\int H\,dt$, the e-folding number in the present framework can be
written as
\begin{eqnarray}
	\label{eq34}
	N
	=
	\int_{\phi_{hc}}^{\phi_{end}}
	\frac{
		d\phi}
	{
		\,a\,\frac{d\phi}{da}
	}\,.
\end{eqnarray}

Here $\phi_{hc}$ denotes the value of the inflaton field at horizon
crossing, while $\phi_{end}$ corresponds to the end of inflation.
Solving Eq.~(\ref{eq34}) determines $\phi_{hc}$ in terms of the
e-folding number and the model parameters. Substituting the result
into Eqs.~(\ref{eq32}) and~(\ref{eq33}) then yields the inflationary
observables $(n_s,r)$ used in the comparison with observational
data.

Constraints on inflationary observables have been reported by
several recent cosmological surveys. The \textit{Planck} 2018
release, combining TT, TE, EE, low-$\ell$ polarization, lensing,
BAO, and BK18 data, gives~\cite{pa22}
$$
r < 0.036
\qquad
(95\%~{\rm CL,\ for}\ \Lambda{\rm CDM}+r+dn_s/d\ln k),
$$
together with the scalar spectral index
\[
n_s=0.9658\pm0.0038.
\]

More recent \textit{DESI} observations
\cite{Ad24,Ad25} further refine the allowed parameter space. In
particular, a joint analysis including BK18 and CMB data
reports~\cite{wa24}
$$
n_s = 0.9700 \pm 0.0036,
\qquad
r = 0.0176^{+0.0070}_{-0.0130}.
$$

Additional constraints obtained from different dataset combinations
within the $\Lambda$CDM framework were presented
in~\cite{co24}. Representative observational bounds used in the
present analysis are summarized in Table~\ref{tab1}. These
constraints provide the observational benchmarks against which the
predictions of the model are compared.

In addition, recent measurements from the Atacama Cosmology
Telescope (ACT) provide complementary constraints on the scalar
spectral index. Combining \textit{Planck} 2018, ACT, lensing, and
BAO (DESI) observations yields~\cite{ACT1}
$$
n_{s}=0.9743 \pm 0.0034\,.
$$

Furthermore, the tensor-to-scalar ratio constrained from a joint
analysis of \textit{Planck} 2018, ACT, and BK18 data gives
~\cite{ACT2}
$$
r < 0.038\,.
$$

\begin{table}[h!]
	\centering
	\begin{tabular}{lcc}
		\hline
		Data Combination & $n_s$ & $r$ \\
		\hline
		SDSS + CMB + Union3
		& $0.9652 \pm 0.0037$
		& $0.0172^{+0.0071}_{-0.013}$ \\
		
		DESI + CMB + Union3
		& $0.9673 \pm 0.0036$
		& $0.0178^{+0.0077}_{-0.013}$ \\
		
		SDSS + CMB + DESY5
		& $0.9644 \pm 0.0036$
		& $0.0169^{+0.0069}_{-0.013}$ \\
		
		DESI + CMB + DESY5
		& $0.9664 \pm 0.0035$
		& $0.0176^{+0.0073}_{-0.013}$ \\
		\hline
	\end{tabular}
	\caption{Representative observational constraints on the scalar
		spectral index $n_s$ and tensor-to-scalar ratio $r$ from
		different dataset combinations~\cite{co24}.}
	\label{tab1}
\end{table}

The constraints summarized above provide the observational
benchmarks against which the predictions of the present
superpotential model are tested. We begin our analysis by examining
the behavior in the $r$--$n_s$ plane using the observational
constraints from \textit{Planck} 2018 TT, TE, EE + lowE + lensing +
BK18 + BAO, together with the DESI + CMB + DESY5 dataset, as well
as the combined \textit{Planck} 2018 + ACT + lensing + BK18 + BAO
(DESI) dataset. The corresponding result is presented in
Fig.~\ref{fig1}.

Figure~\ref{fig1} illustrates how variations of the parameters
$\beta$ and $C$ modify the theoretical trajectory in the
$r$--$n_s$ plane, allowing a direct comparison with the
$68\%$ and $95\%$ confidence contours from the observational data.
For the numerical analysis, we consider the parameter ranges
$0<C<1$
and
$-5\times10^{2}\leq\beta\leq0$. As indicated by the numerical analysis, the observationally viable regions are predominantly concentrated in the negative-$\beta$ sector. This preference reflects the impact of the GUP-induced corrections on the inflationary observables and the resulting displacement of the theoretical trajectory in the $(n_s,r)$ plane. Consequently, moderately negative values of $\beta$ shift the model predictions toward the observationally preferred region. As shown in Fig.~\ref{fig1}, the model predictions become compatible with current observational constraints for specific
regions of the parameter space. In particular, part of the
theoretical region overlaps with the confidence contours obtained
from the \textit{Planck} 2018 and DESI datasets, indicating
consistency with present cosmological observations.

Based on our analysis, the $\phi^{2}$-superpotential inflationary
model with GUP-induced corrections is consistent with the
\textit{Planck} 2018 TT, TE, EE + lowE + lensing + BK18 + BAO data
at the $95\%$ confidence level for
$-315\leq\beta\leq-1.00$, depending on the value of
$0<C\leq0.556$. The model is compatible with the same dataset at the
$68\%$ confidence level for
$-299\leq\beta\leq-8.63$, depending on the value of
$0<C\leq0.548$.

Using the DESI + CMB + DESY5 dataset, the model remains consistent
with observations at the $95\%$ confidence level for
$-315\leq\beta\leq-5.00$, depending on the value of
$0.28<C\leq0.559$. At the $68\%$ confidence level, consistency is
obtained for
$-298\leq\beta\leq-26.5$, depending on the value of
$0.28<C\leq0.555$.

Our numerical analysis further shows that the model is compatible
with the combined \textit{Planck} 2018 + ACT + lensing + BK18 + BAO
(DESI) dataset at the $95\%$ confidence level for
$-319\leq\beta\leq-6.67$, depending on the value of
$0.32<C\leq0.556$. The corresponding $68\%$ confidence region is
obtained for
$-299\leq\beta\leq-14.0$, depending on the value of
$0.32<C\leq0.556$.

Figures~\ref{fig2}--\ref{fig4} illustrate the allowed regions in the
$(\beta,C)$ parameter space for which the model yields values of the
scalar spectral index and tensor-to-scalar ratio compatible with
observations. Figure~\ref{fig2} corresponds to the
\textit{Planck} 2018 TT, TE, EE + lowE + lensing + BK18 + BAO
dataset, while Fig.~\ref{fig3} is based on the DESI + CMB + DESY5
dataset. Figure~\ref{fig4} presents the results obtained using the
combined \textit{Planck} 2018 + ACT + lensing + BK18 + BAO (DESI)
dataset. All figures display the $68\%$ and $95\%$ confidence
regions.

We have also derived constraints on the parameter $\beta$ for
selected values of $C$ using the observational datasets discussed
above. The corresponding numerical results are summarized in
Table~\ref{tab2}.

\begin{figure}[htbp]
	\centering
	\includegraphics[width=0.65\textwidth]{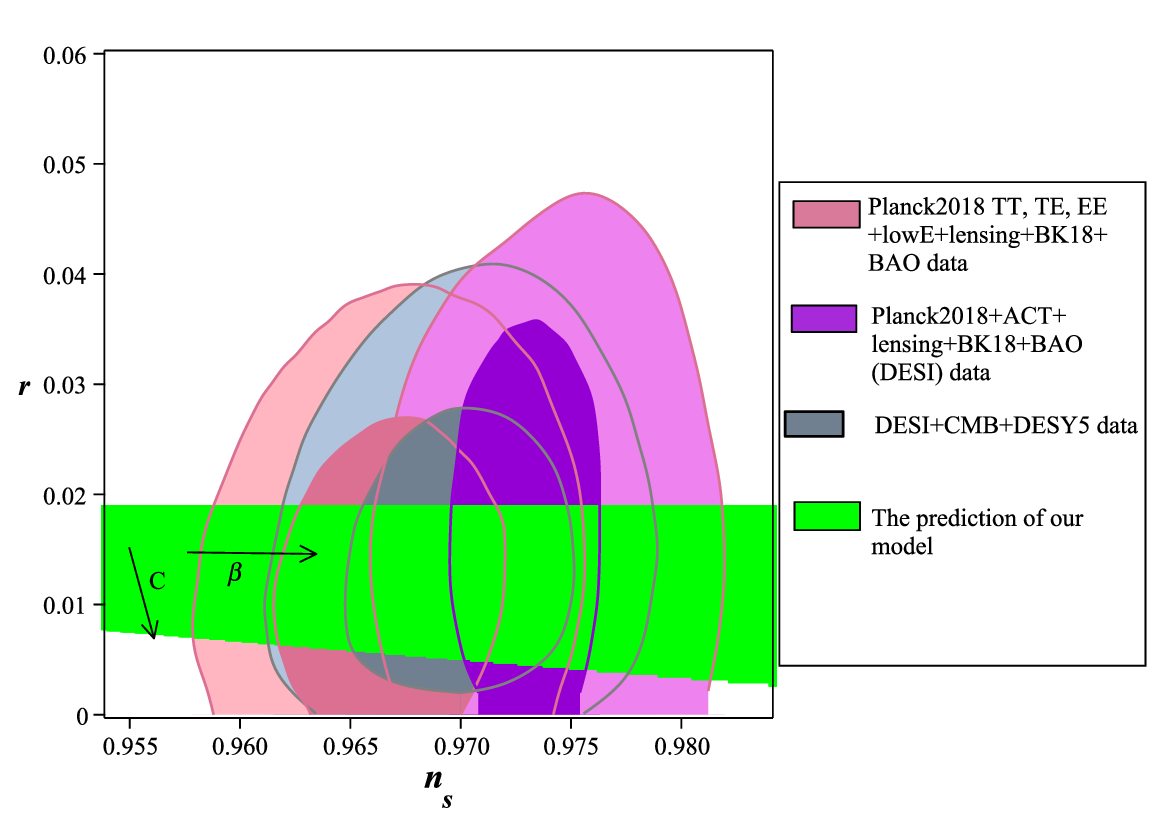}
	\caption{\small
		Tensor-to-scalar ratio versus scalar spectral index for the
		$\phi^{2}$-superpotential inflationary model with GUP-induced
		corrections. The comparison is performed using the
		\textit{Planck} 2018 TT, TE, EE + lowE + lensing + BK18 + BAO,
		DESI + CMB + DESY5, and
		\textit{Planck} 2018 + ACT + lensing + BK18 + BAO (DESI)
		datasets, with the number of e-folds fixed at $N=60$.
		The arrows indicate the directions of increasing model parameters.
	}
	\label{fig1}
\end{figure}

\begin{figure}[htbp]
	\centering
	\includegraphics[width=0.45\textwidth]{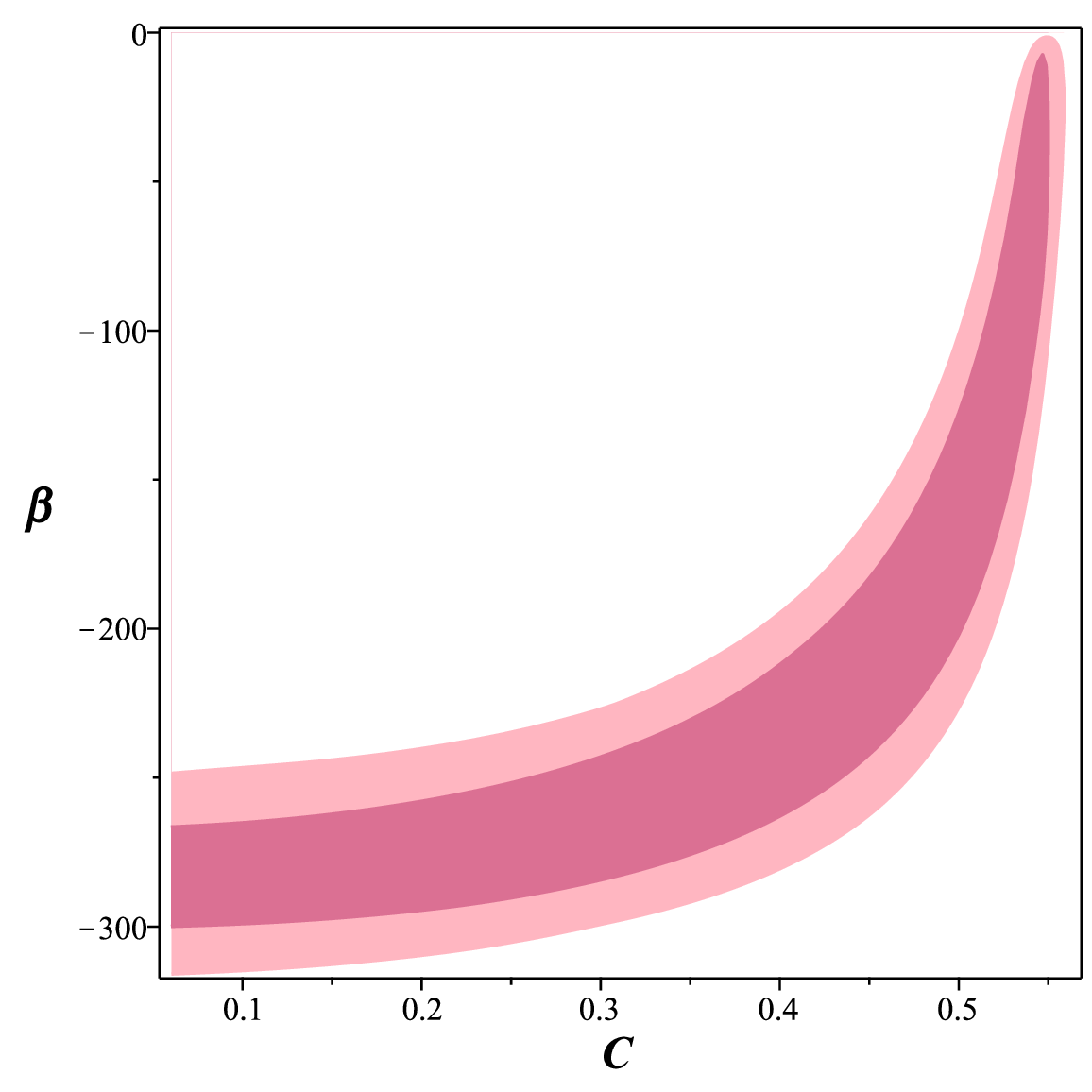}
	\caption{\small
		Allowed regions in the $(C,\beta)$ parameter space for which the
		$\phi^{2}$-superpotential inflationary model with GUP-induced
		corrections yields values of $(n_s,r)$ compatible with
		observations. The figure is constructed using the
		\textit{Planck} 2018 TT, TE, EE + lowE + lensing + BK18 + BAO
		dataset at the $68\%$ and $95\%$ confidence levels.
	}
	\label{fig2}
\end{figure}

\begin{figure}[htbp]
	\centering
	\includegraphics[width=0.45\textwidth]{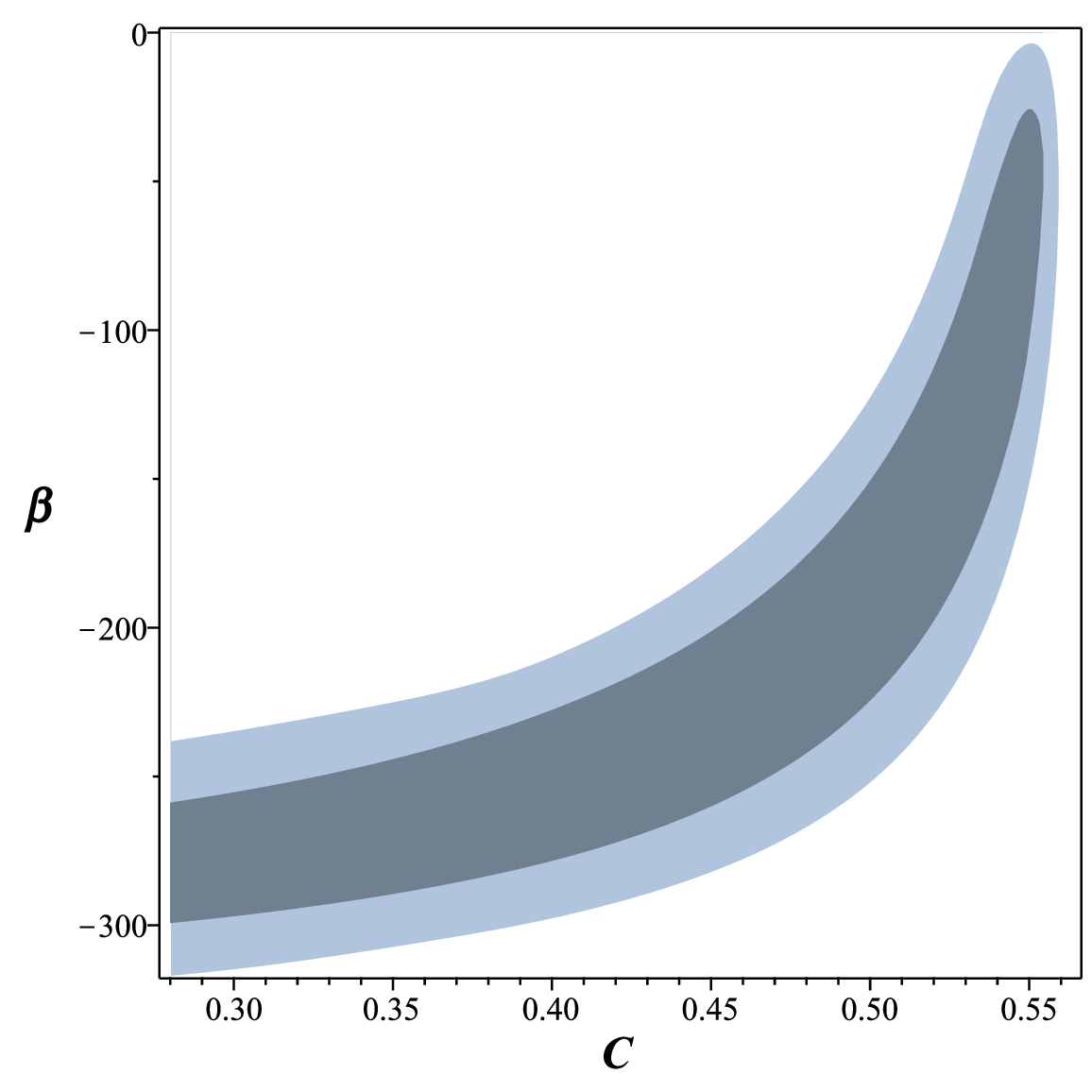}
	\caption{\small
		Allowed regions in the $(C,\beta)$ parameter space for which the
		$\phi^{2}$-superpotential inflationary model with GUP-induced
		corrections yields values of $(n_s,r)$ compatible with
		observations. The figure is constructed using the
		DESI + CMB + DESY5 dataset at the $68\%$ and $95\%$
		confidence levels.
	}
	\label{fig3}
\end{figure}

\begin{figure}[htbp]
	\centering
	\includegraphics[width=0.45\textwidth]{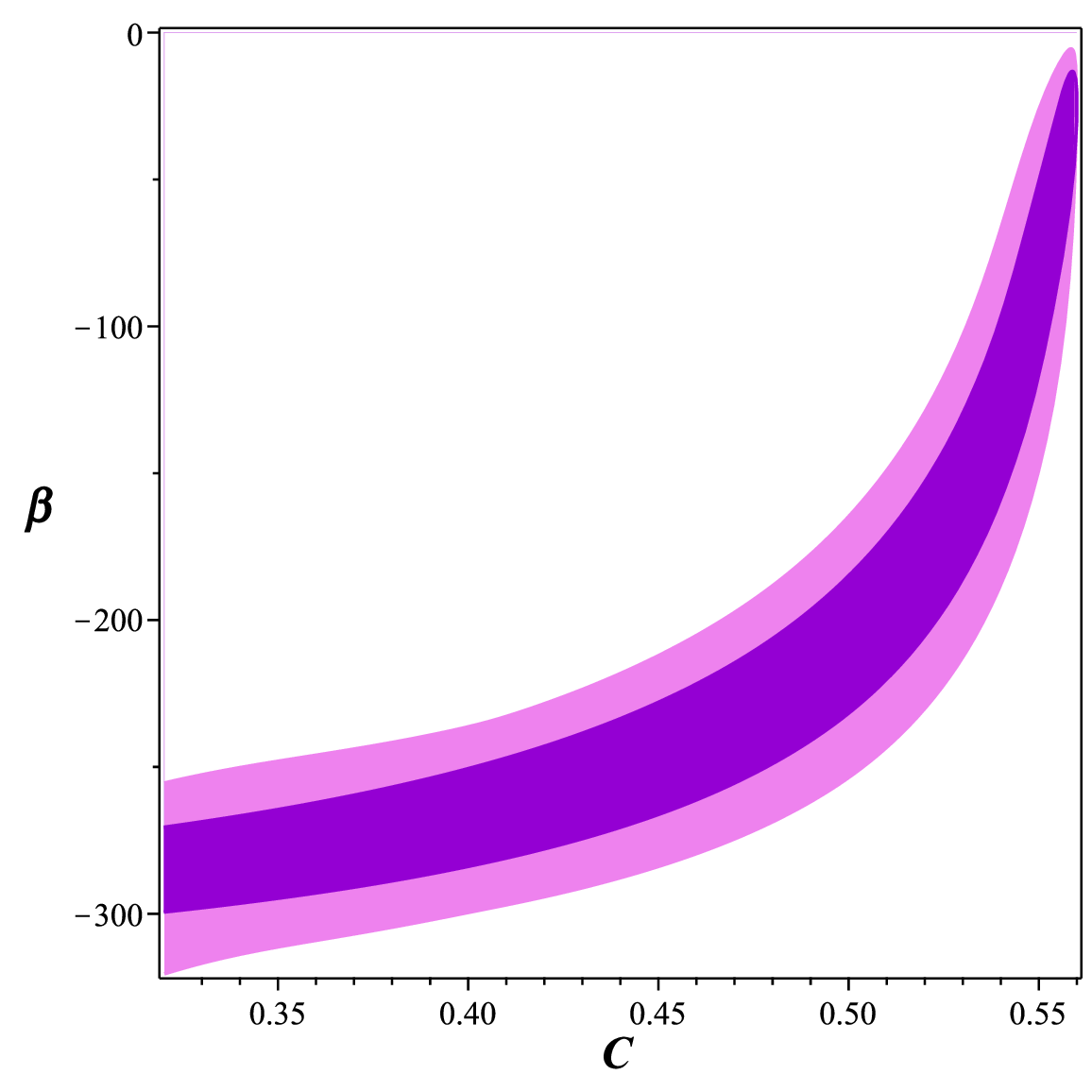}
	\caption{\small
		Allowed regions in the $(C,\beta)$ parameter space for which the
		$\phi^{2}$-superpotential inflationary model with GUP-induced
		corrections yields values of $(n_s,r)$ compatible with
		observations. The figure is constructed using the
		combined \textit{Planck} 2018 + ACT + lensing + BK18 + BAO
		(DESI) dataset at the $68\%$ and $95\%$ confidence levels.
	}
	\label{fig4}
\end{figure}

\begin{figure}[htbp]
	\centering
	\includegraphics[width=0.45\textwidth]{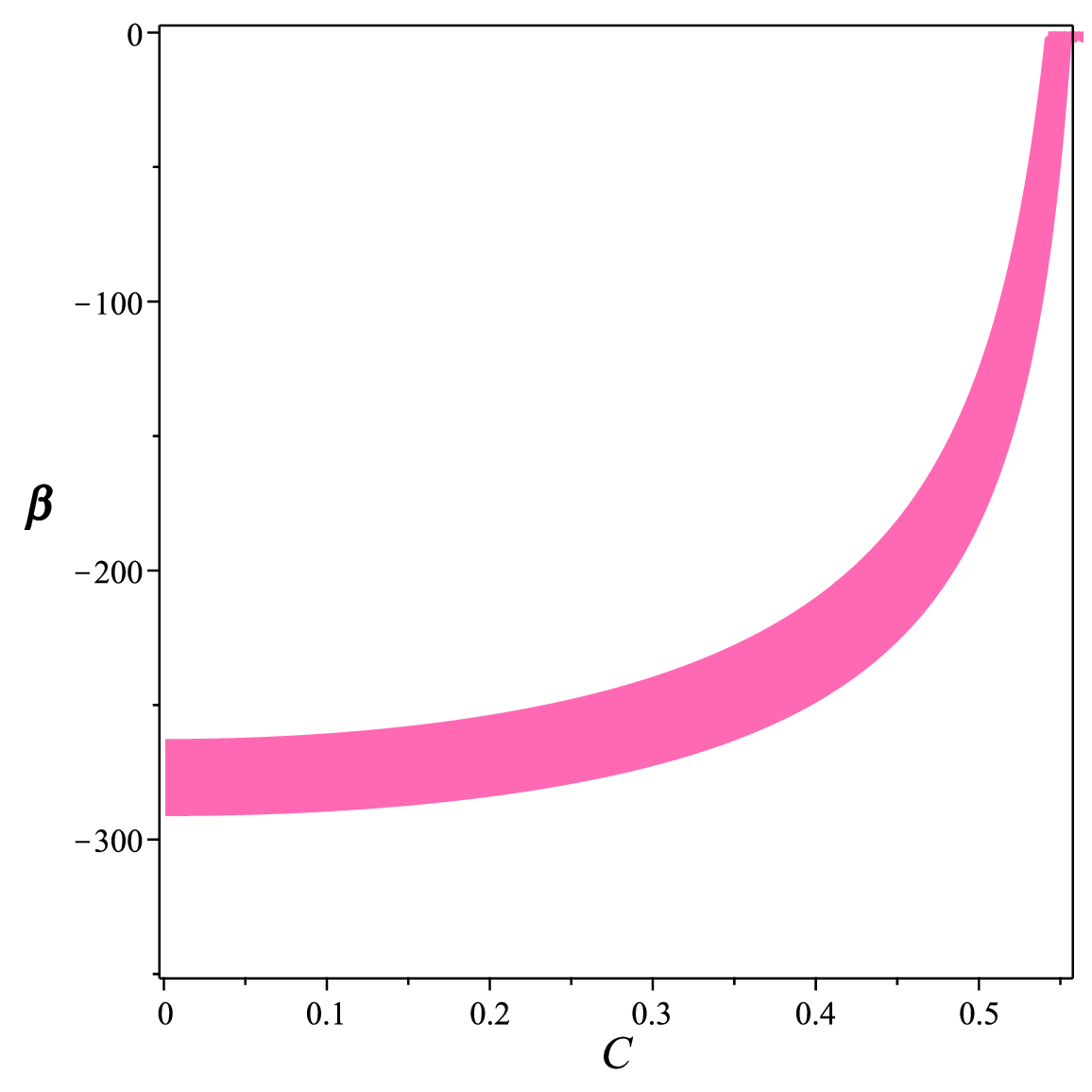}
	\includegraphics[width=0.45\textwidth]{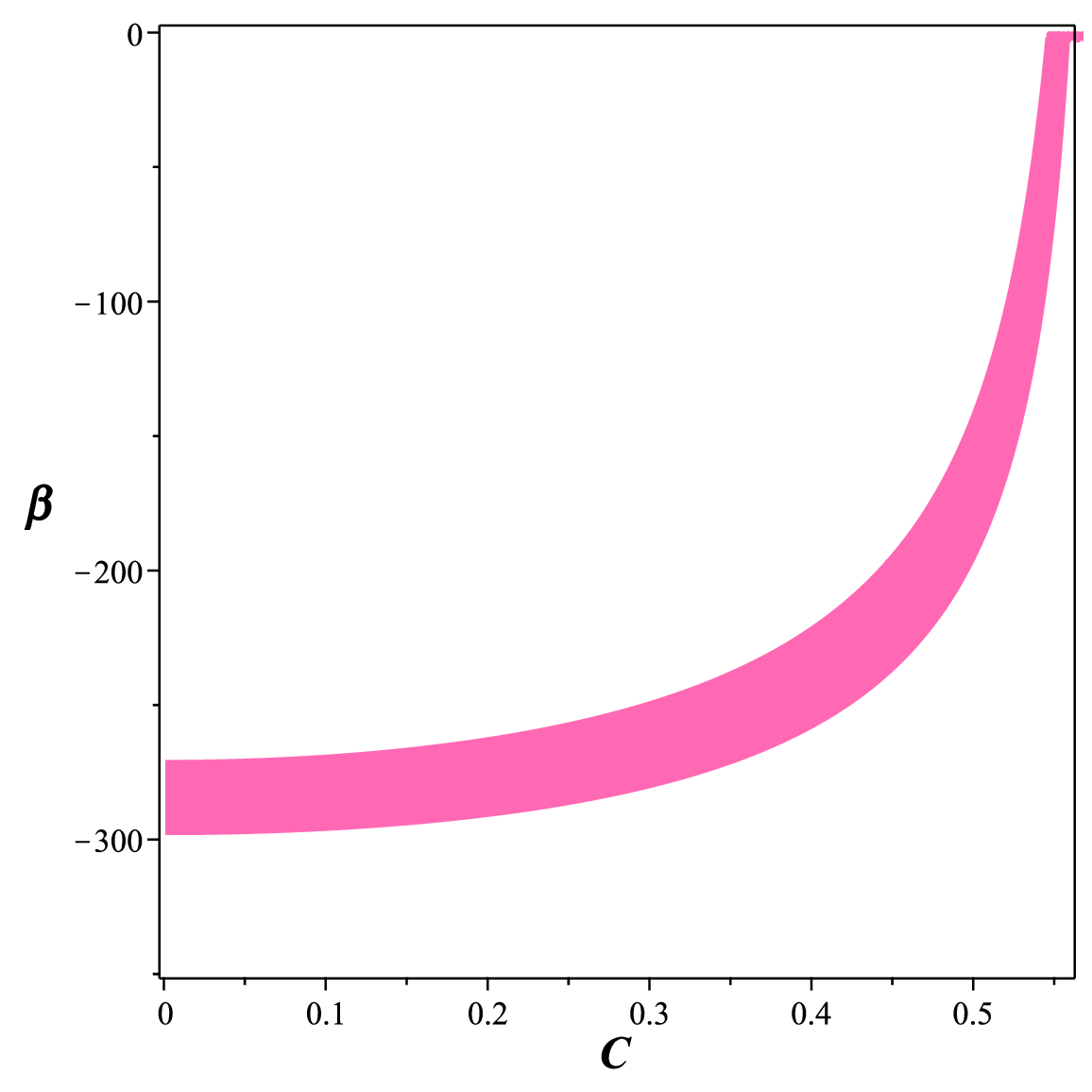}
	
	\includegraphics[width=0.45\textwidth]{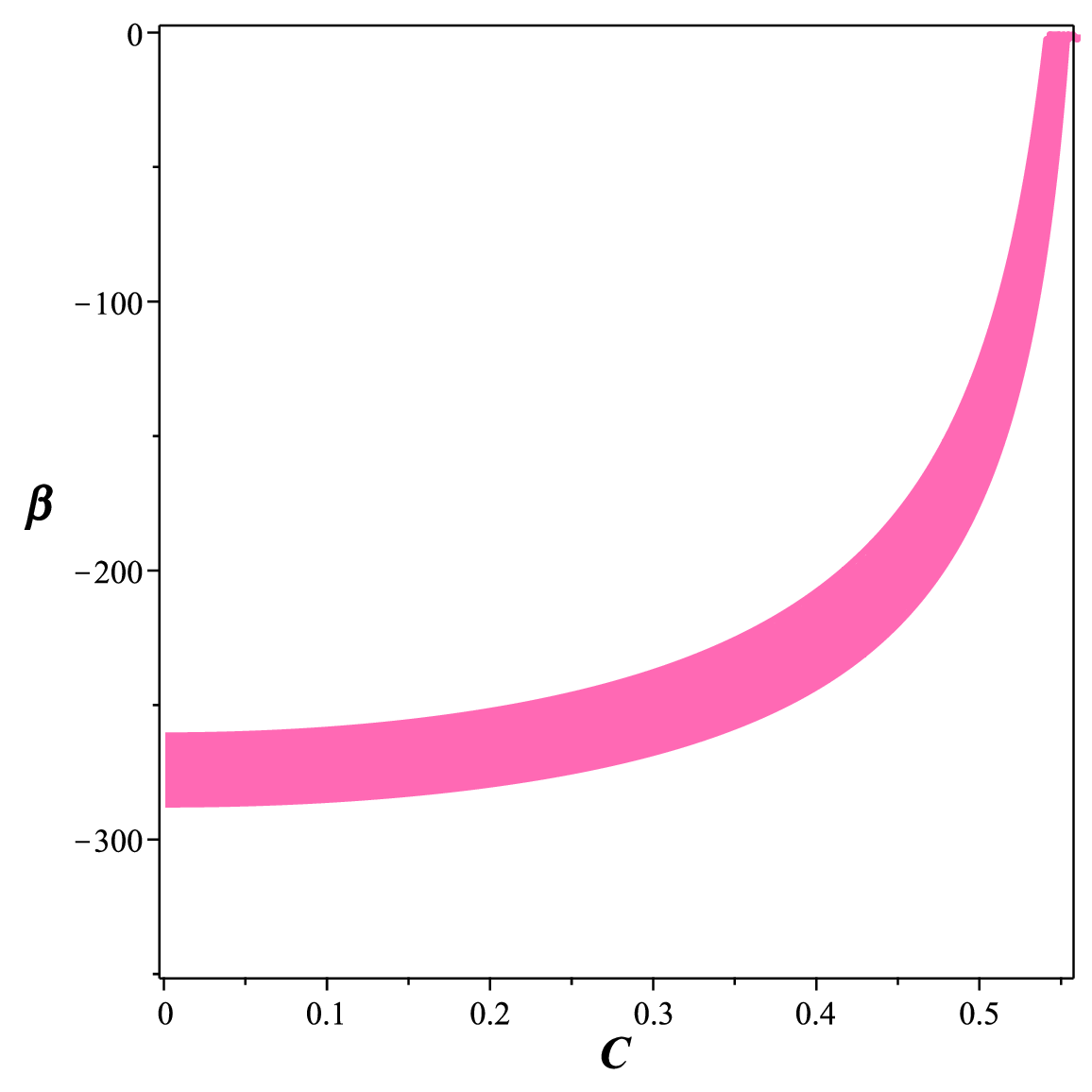}
	\includegraphics[width=0.45\textwidth]{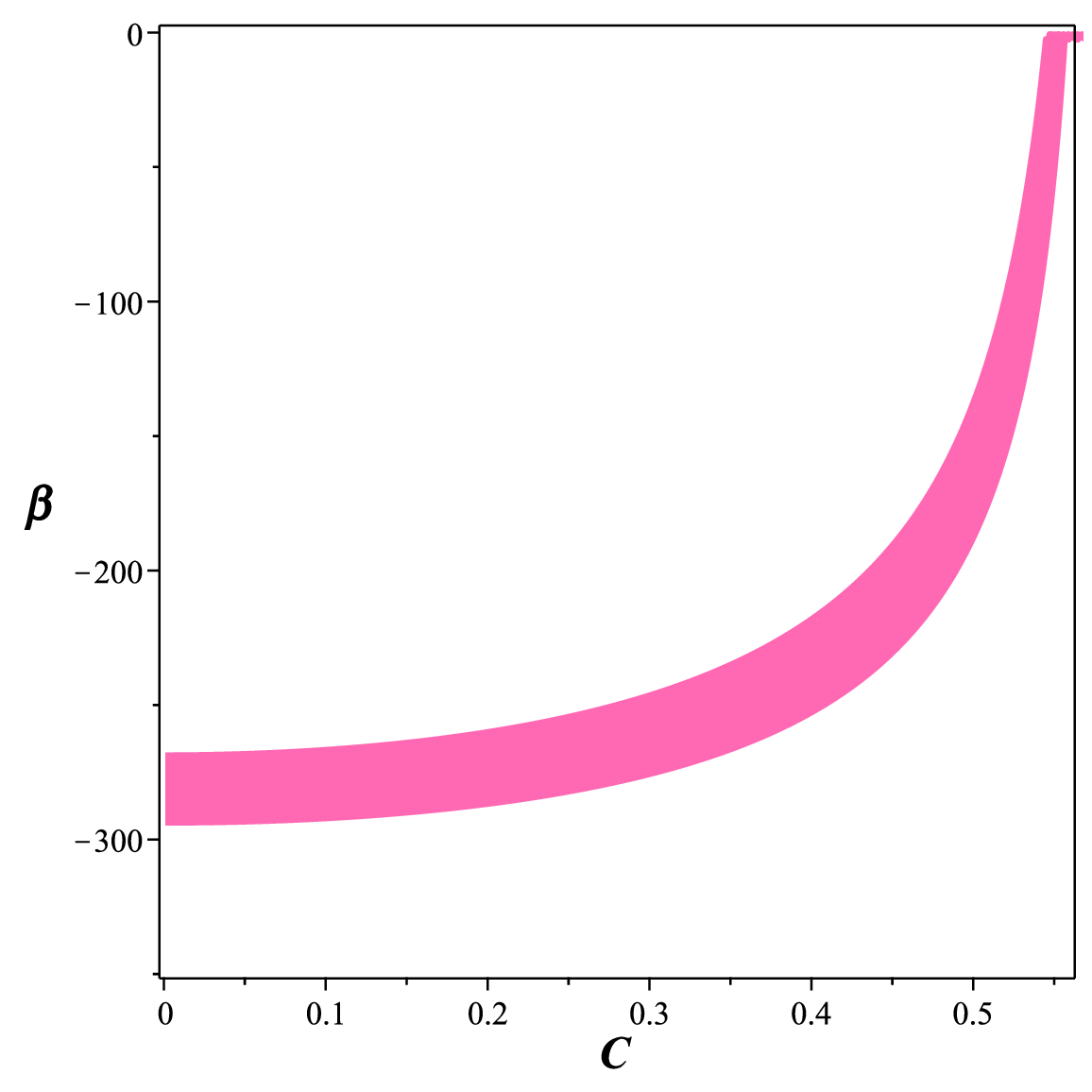}
	
	\caption{\small
		Allowed regions in the $(\beta,C)$ parameter space for which the
		$\phi^{2}$-superpotential inflationary model with GUP-induced
		corrections reproduces values of $(n_s,r)$ compatible with
		observational constraints. The panels correspond respectively to
		SDSS + CMB + Union3 (upper left),
		DESI + CMB + Union3 (upper right),
		SDSS + CMB + DESY5 (lower left), and
		DESI + CMB + DESY5 (lower right).
		The shaded regions indicate parameter combinations compatible
		with each dataset.
	}
	\label{fig5}
\end{figure}

\begin{table*}[htbp]
	\centering
	\scriptsize
	\tabcolsep=2pt
	\renewcommand{\arraystretch}{1.15}
	
	\caption{\small
		Allowed ranges of the parameter $\beta$ for selected values of
		$C$, for which the $\phi^{2}$-superpotential inflationary model
		with GUP-induced corrections yields values of $(n_s,r)$
		compatible with different observational datasets at the
		$68\%$ and $95\%$ confidence levels.
	}
	\label{tab2}
	
	\begin{tabular}{|c|c|c|c|c|c|c|}
		\hline
		
		&
		\multicolumn{2}{c|}{
			\shortstack{
				\textit{Planck} 2018 \\
				TT, TE, EE + lowE \\
				+lensing + BK18 + BAO
			}
		}
		&
		\multicolumn{2}{c|}{
			\shortstack{
				DESI + CMB \\
				+ DESY5
			}
		}
		&
		\multicolumn{2}{c|}{
			\shortstack{
				\textit{Planck} 2018 + ACT \\
				+lensing + BK18 \\
				+ BAO (DESI)
			}
		}
		\\
		
		\cline{2-7}
		
		$C$
		& $68\%$ CL
		& $95\%$ CL
		& $68\%$ CL
		& $95\%$ CL
		& $68\%$ CL
		& $95\%$ CL
		\\
		
		\hline\hline
		
		$0.2$
		& $-295<\beta<-258$
		& $-310<\beta<-240$
		& ---
		& ---
		& ---
		& ---
		\\
		
		\hline
		
		$0.3$
		& $-285<\beta<-242$
		& $-299<\beta<-226$
		& $-296<\beta<-255$
		& $-313<\beta<-235$
		& ---
		& ---
		\\
		
		\hline
		
		$0.4$
		& $-263<\beta<-210$
		& $-280<\beta<-194$
		& $-278<\beta<-228$
		& $-296<\beta<-209$
		& $-282<\beta<-250$
		& $-299<\beta<-235$
		\\
		
		\hline
		
		$0.5$
		& $-204<\beta<-131$
		& $-226<\beta<-98.7$
		& $-222<\beta<-152$
		& $-250<\beta<-123$
		& $-230<\beta<-183$
		& $-254<\beta<-163$
		\\
		
		\hline
	\end{tabular}
\end{table*}

To complement our analysis, we further examine the observationally
viable regions of the parameter space spanned by $\beta$ and $C$.
Figure~\ref{fig5} displays the corresponding allowed regions that
produce values of $(n_s,r)$ compatible with different observational
datasets. The upper-left panel corresponds to
SDSS + CMB + Union3, the upper-right to
DESI + CMB + Union3, the lower-left to
SDSS + CMB + DESY5, and the lower-right to
DESI + CMB + DESY5. The shaded regions indicate the sets of
$(\beta,C)$ values yielding predictions compatible with the
corresponding observational constraints on the scalar spectral index
and tensor-to-scalar ratio. The clustering of the viable regions toward negative values of $\beta$ indicates that the GUP-induced corrections shift the model predictions toward observationally favored regions characterized by
smaller values of $r$ and slightly larger values of $n_s$.
We have also derived constraints on the parameter $\beta$ for
selected values of $C$ using these datasets. The corresponding
results are summarized in Table~\ref{tab3}.

Overall, the comparison across multiple datasets shows that the
$\phi^{2}$-superpotential inflationary model remains compatible with
current observations for moderately negative values of $\beta$ and
sub-unity values of $C$. These results suggest that GUP-induced
corrections can significantly modify the inflationary predictions of
the model and improve their agreement with high-precision
cosmological observations.

\begin{table*}[htbp]
	\centering
	\scriptsize
	\renewcommand{\arraystretch}{1.2}
	
	\caption{\small
		Allowed ranges of the parameter $\beta$ for selected values of
		$C$, for which the $\phi^{2}$-superpotential inflationary model
		with GUP-induced corrections yields values of $(n_s,r)$
		compatible with different observational datasets at the
		$68\%$ confidence level.
	}
	\label{tab3}
	
	\begin{tabular}{|c|c|c|c|}
		\hline\hline
		
		$C$
		& SDSS + CMB + Union3
		& DESI + CMB + Union3
		& SDSS + CMB + DESY5
		\\
		
		\hline\hline
		
		$0.2$
		& $-282<\beta<-255$
		& $-290<\beta<-261$
		& $-280<\beta<-250$
		\\
		
		\hline
		
		$0.3$
		& $-271<\beta<-241$
		& $-280<\beta<-248$
		& $-267<\beta<-237$
		\\
		
		\hline
		
		$0.4$
		& $-248<\beta<-211$
		& $-256<\beta<-221$
		& $-245<\beta<-207$
		\\
		
		\hline
		
		$0.5$
		& $-180<\beta<-130$
		& $-194<\beta<-141$
		& $-175<\beta<-120$
		\\
		
		\hline
	\end{tabular}
\end{table*}

\subsection{A Short Comparison with Leading Inflationary Scenarios}

It is instructive to compare the present framework with several leading inflationary scenarios. In canonical quadratic inflation~\cite{Linde1983}, the main tension with current observations originates from the relatively large predicted tensor-to-scalar ratio~\cite{pl18b}. By contrast, the GUP-corrected superpotential model studied here shifts the theoretical trajectory toward smaller values of $r$, allowing substantial regions of the parameter space to enter the observationally favored domain. In this sense, the improvement does not arise from a flattening of the scalar potential, but rather from modifications of the background dynamics and perturbation sector induced by the GUP-corrected Friedmann equations and the corresponding deformation of the effective comoving wave number.

A different route toward observational viability is realized in Starobinsky inflation and related $R^{2}$ models~\cite{Starobinsky1980}, where the agreement with observations originates from modifications of the gravitational sector~\cite{pl18b,ACT2}. Likewise, $\alpha$-attractor models~\cite{Kallosh2013a,Kallosh2013b} achieve robust predictions through an attractor mechanism that largely suppresses sensitivity to the microscopic details of the theory~\cite{pl18b}. The present framework differs conceptually from both classes of models. Here, the shift toward the observationally preferred region is generated by quantum-gravity-inspired corrections entering through horizon thermodynamics and the generalized uncertainty principle. Consequently, the viable parameter regions retain sensitivity to the deformation parameter $\beta$ and the superpotential parameter $C$. Unlike attractor-type models, whose predictions become largely insensitive to the underlying microscopic parameters, the present framework allows cosmological observations to place direct constraints on quantities associated with quantum-gravity corrections. In this respect, the model provides an alternative phenomenological route toward observational viability while preserving sensitivity to the parameters controlling the underlying quantum-gravity effects.

The robustness and statistical relevance of the viable parameter regions are supported by the consistency of the results obtained from different observational datasets. Although the precise boundaries of the allowed regions vary slightly among the Planck, DESI, and ACT analyses, both the $68\%$ and $95\%$ confidence regions consistently favor moderately negative values of $\beta$ together with sub-unity values of $C$. Moreover, the substantial overlap between the parameter intervals inferred from these independent datasets indicates that the observational viability of the model is not associated with an isolated contour crossing in the $(n_s,r)$ plane or with a narrowly tuned parameter choice. Rather, it reflects a stable feature of the parameter space within the present framework. At the same time, we emphasize that the present analysis is based on confidence-region comparisons in the $(n_s,r)$ plane rather than on a full Bayesian model-selection study. Therefore, the results should be interpreted as evidence for the observational viability and stability of the identified parameter regions, rather than as a statistical preference over alternative inflationary scenarios.

\section{Summary and Conclusions}

In this work, we investigated a
$\phi^{2}$-superpotential inflationary scenario within a
GUP-corrected framework and examined its consistency with current
cosmological observations. Starting from the first-law formulation
of horizon thermodynamics, we derived modified Friedmann equations
by combining the temperature--surface-gravity relation with a
generalized entropy-area law. The resulting GUP-corrected
background equations, Eqs.~(\ref{eq17})--(\ref{eq18}), were then
used as the basis for the subsequent inflationary analysis. To connect the modified background dynamics with the
superpotential formalism, we treated the inflaton field as
$\phi(a)$ and reformulated the dynamics in terms of the
superpotential $W(\phi)$ and its derivatives. This allowed us to
construct the corresponding slow-roll parameters in the
GUP-corrected setup. On the perturbation side, we adopted an
effective implementation of the GUP-deformed commutation
relations, leading to a modified comoving wave number and
corresponding corrections to the scalar spectral index, tensor
spectral index, and tensor-to-scalar ratio. These expressions
demonstrate how GUP-induced corrections modify the standard
inflationary consistency relation.

For the phenomenological analysis, we adopted the quadratic
superpotential ansatz
\[
W(\phi)=\tfrac{1}{2}m^{2}\phi^{2},
\]
and evaluated the inflationary observables $n_{s}$ and $r$ as
functions of the model parameters. Confronting the predictions
with the \textit{Planck} 2018 (TT, TE, EE + lowE + lensing +
BK18 + BAO), DESI + CMB + DESY5, and
\textit{Planck} 2018 + ACT + lensing + BK18 + BAO (DESI)
datasets, we mapped the viable regions in the $(\beta,C)$
parameter space and summarized representative observational
constraints in Table~\ref{tab1}. The comparison in the
$r$--$n_s$ plane (Fig.~\ref{fig1}, with $N=60$) shows partial
overlap between the model predictions and the $68\%$ and
$95\%$ confidence contours for specific parameter intervals.
Detailed parameter-space scans
(Figs.~\ref{fig2}--\ref{fig5}) further indicate that
compatibility with current observations is achieved for
moderately negative values of $\beta$ and sub-unity values of
$C$. Explicit allowed ranges of $\beta$ for selected values of
$C$ obtained from different dataset combinations were presented
in Tables~\ref{tab2} and~\ref{tab3}. Overall, the results suggest that GUP-induced corrections can
significantly modify the $(n_s,r)$ predictions of the quadratic
superpotential model and improve their agreement with current
high-precision cosmological observations.

The thermodynamic derivation of the GUP-corrected background
equations, their reformulation within the superpotential
framework, and the effective perturbation-level modifications
induced by the GUP deformation together lead to a
self-consistent inflationary framework in which the
$\phi^{2}$-superpotential model remains compatible with current
observations for
$0<C\!\lesssim\!0.56$
and
$-319\!\lesssim\!\beta<0$
within the parameter ranges favored by the \textit{Planck} and
DESI datasets. The analysis highlights the potential role of
quantum-gravity motivated corrections in improving the agreement
between simple superpotential inflationary constructions and
high-precision cosmological observations. The resulting shift in
the $(n_s,r)$ plane does not arise merely from an effective
rescaling of slow-roll trajectories, but is instead associated
with the modified comoving wave number induced by the GUP
deformation, thereby modifying the standard inflationary
consistency relation.

\textbf{ACKNOWLEDGMENTS}\\
We thank the referees for the very insightful comments that have	improved the quality of the paper considerably.
\\

\textbf{Data Availability Statement:} All relevant analytical expressions and numerical results are included in the manuscript.

\textbf{Code/Software Availability Statement:}
No public code repository is associated with this work.

\end{document}